\documentclass[12pt,letterpaper,english]{iopart}
\usepackage[T1]{fontenc}
\usepackage[latin1]{inputenc}
\usepackage{prettyref}
\usepackage{graphicx}

\makeatletter

\usepackage{babel}
\makeatother
\begin{document}

\title{Monolithic Geometric Anti-Spring Blades}

\author{G Cella\dag, V Sannibale\ddag
\footnote[8]{Corresponding author. Address: California Institute of Technology,
LIGO Project, 1200 E. California Blvd. 91125, Pasadena, CA USA, MS
18-34 phone +626-395-6358\\ LIGO DCC number: P040004-00-D
}, R DeSalvo\ddag, S M\'arka\ddag, A Takamori\S}

\address{\dag Universit\`a di Pisa, Pisa, Italy}

\address{\ddag California Institute of Technology, LIGO Project, Pasadena,
California}

\address{\S University of Tokyo, Tokyo, Japan}

\ead{sannibale\_v@ligo.caltech.edu}

\begin{abstract}
In this article we investigate the principle and properties of a vertical
passive seismic noise attenuator conceived for ground based gravitational
wave interferometers. This mechanical attenuator based on a particular
geometry of cantilever blades called monolithic geometric anti springs
(MGAS) permits the design of mechanical harmonic oscillators with very
low resonant frequency (below 100mHz).

Here we address the theoretical description of the mechanical device,
focusing on the most important quantities for the low frequency regime,
on the distribution of internal stresses, and on the thermal stability.

In order to obtain physical insight of the attenuator peculiarities,
we devise some simplified models, rather than use the brute force
of finite element analysis. Those models have been used to optimize
the design of a seismic attenuation system prototype for LIGO advanced
configurations and for the next generation of the TAMA interferometer.

\end{abstract}

\section{Introduction}

A fundamental issue in designing interferometric gravitational wave
detectors is the isolation of the test masses (heavy mirrors) from
environmental noise sources, such as seismic noise. It is well known
that a very effective way to reduce this type of perturbation is to
use mechanical harmonic oscillators, which possess the important feature
of attenuating the seismic noise above their resonant frequency.

One of the major difficulties in designing this kind of system is
to achieve vertical attenuation performance comparable to that obtainable
in the horizontal direction. In long baseline interferometers, this
is a significant issue because the vertical motion is coupled to the
direction of sensitivity (horizontal direction) through mechanical
imperfections, optic wedges, the earth curvature, and mirrors' orientation.

A possible solution to the problem, which has been implemented in
the so-called super-attenuators of the VIRGO detector {[}Caron \etal
1997{]}, is based on the magnetic anti-spring concept {[}Beccaria
\etal 1997{]}.

A considerable improvement to the magnetic anti-spring is the geometric
anti-spring concept (GAS) {[}Bertolini \etal 1999, Cella \etal 2002{]},
which allows significant increases in the vertical attenuation and
thermal stability of a seismic attenuation system (seismic
filter).

The monolithic geometric anti-spring (MGAS) concept introduced by one
of the authors is essentially an evolution of the GAS concept capable
to significantly improve the GAS system performance.

A seismic attenuation system based on the MGAS concept is better than
the equivalent GAS system mainly because of its simplicity. In fact,
this characteristic allows removal of several mechanical parts that
introduce unwanted low frequency resonances responsible for reducing
the filtering efficiency on all the degrees of freedom.

In the next sections we address the theoretical description of the
MGAS, and some of the main issues related to the design of a seismic
filter.

\section{Low Frequency One Dimensional Model}

\begin{figure}
\begin{center}\includegraphics[%
%bb = 0 0 200 100, draft, type=eps]{fig1.eps}
width=1.0\linewidth]{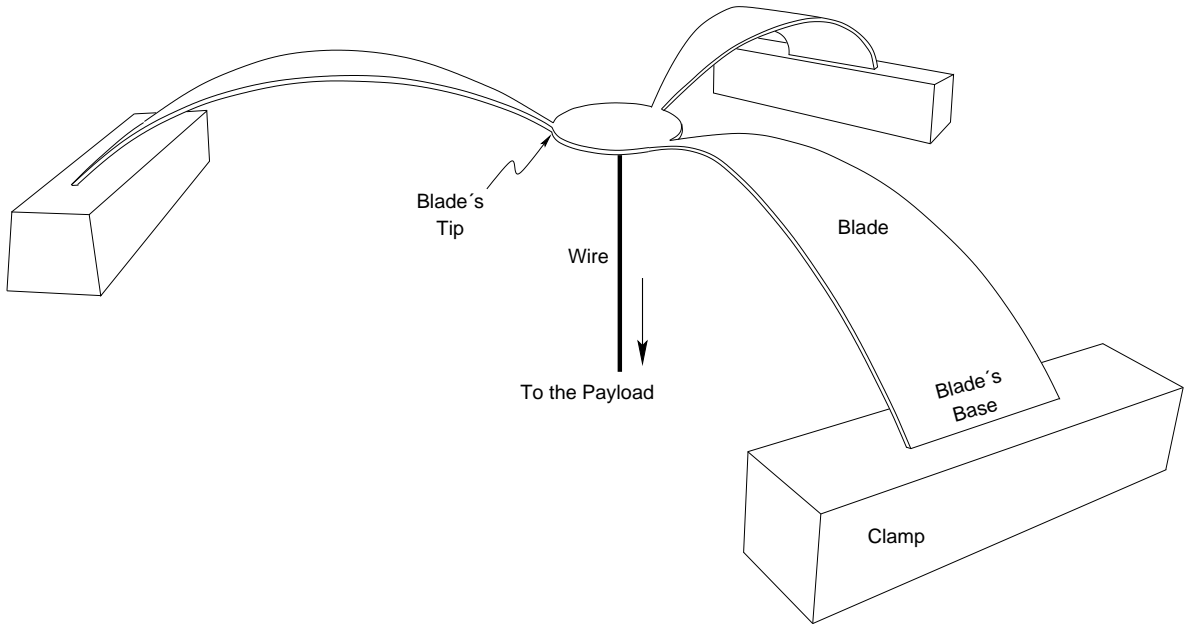}
\end{center}

\caption{\label{cap:MGASF-scheme}Schematic view of a complete MGAS filter.}
\end{figure}

An MGAS configuration is made of several quasi-triangular blades radially
disposed and connected together at their vertices (see Figure~\ref{cap:MGASF-scheme}).

\begin{figure}[htbp]
\begin{center}\includegraphics[%
  width=1.0\linewidth]{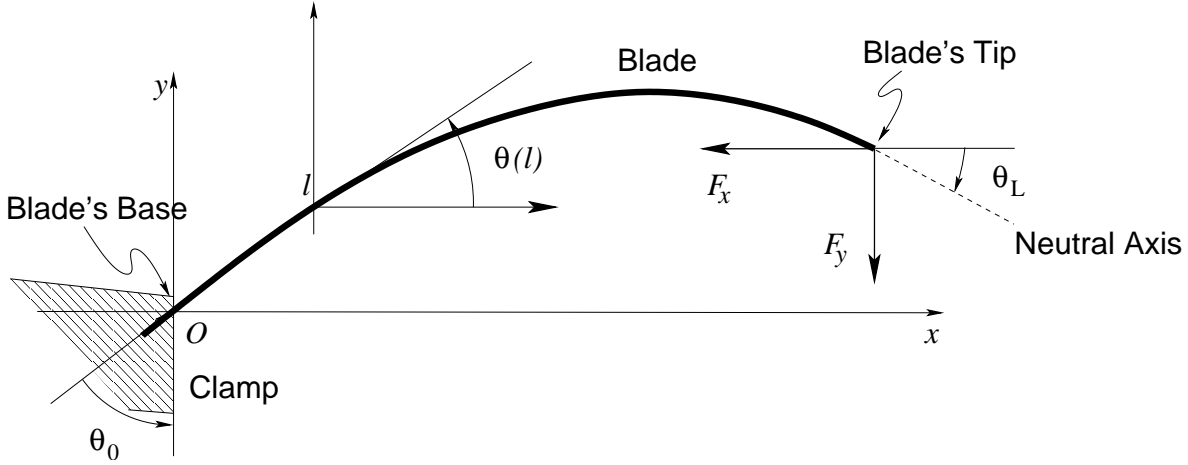}\end{center}

\caption{\label{cap:Monolithic-geometric-anti-spring}Reference Frame for
the MGAS study. The curvilinear coordinate $l$ is measured from the
point O and the component $F_{x}$, and $F_{y}$ represent the external
forces applied to the blade}
\end{figure}

Taking advantage of this radial symmetry, the MGAS physics can be
studied considering just a single blade. Moreover, because of the
blade's symmetry along a vertical plane (called $Oxy$ as shown Figure~\ref{cap:Monolithic-geometric-anti-spring}),
and the typical blade dimensions (thin blade approximation), we can
reduce the dynamics to one of a simple single degree of freedom system.

The blade's base is clamped to a massive structure and subjected to
an external force on the tip due to the load and to the other radially
arranged blades. The horizontal component compresses and bends the
blade longitudinally.

In the low frequency regime, below the internal mode frequencies,
the dynamics is dominated by the clamping structure mass (much greater
than the blade's mass, which we can neglect) and by the load.

The blade has a constant thickness $d$ much smaller than the two
other dimensions, a length $L$ and a variable transverse width $w(l)$,
with $l\in[0,L]$. The variable $l$ is the curvilinear coordinate
following the intersection of the blade's neutral surface with the
plane $Oxy$, i.e. the neutral axis of the blade. It is also convenient
to define $\theta(l)$ as the angle between the tangent of the curvilinear
coordinate and the $x$ axis.

Under these approximations, we can model the thin blade as a massless
elastic line with a potential energy {[}Landau, Lifshitz 1959{]}
\begin{equation}
U=\int_{0}^{L}\left\{ \frac{1}{2}EI(l)\left[\frac{d\theta(l)}{dl}\right]^{2}-F_{x}\cos\theta(l)-F_{y}\sin\theta(l)\right\} dl.\label{eq:potential}
\end{equation}
$F_{x}$ and $F_{y}$, are respectively the horizontal and vertical
external forces applied to the blade's tip, and $I$ is the transverse
moment of inertia of the blade, $I=w(l)d^{3}/12$. The mechanical
characteristics of the blade material enter just through the Young's
modulus $E$.

It is convenient to rewrite the potential energy in the following
way\begin{equation}
U=\frac{Ed^{3}w(0)}{12L}\tilde{U},\label{eq:potentialAd}\end{equation}
where\begin{equation}
\tilde{U}=\int_{0}^{1}\left\{ \frac{1}{2\gamma(\xi)}\left[\frac{d\theta(\xi)}{d\xi}\right]^{2}-G_{x}\:\cos\theta(\xi)-G_{y}\:\sin\theta(\xi)\right\} d\xi,\label{eq:adimUdef}\end{equation}
and $\gamma(x)=w(0)/w(x)$ is a normalized blade's shape function,
$\xi=l/L$. The quantities $G_{x},G_{y}$ are dimensionless parameters
proportional to the horizontal compression and to the vertical load:\begin{equation}
G_{i}=\frac{12L^{2}}{Ed^{3}w(0)}F_{i},\qquad i=x,\: y\,.\label{eq:Gi}\end{equation}

Writing $U$ in such a form we make clear that the nonlinear behavior
of the blade (the $\tilde{U}$ term) is parameterized by the dimensionless
parameters $G_{i}$, which completely defines the {}``working point''
of the system for a given normalized shape of the
blade $\gamma(\xi)$. Every physical quantity we are interested in
will be indeed written as a product of two terms. The first is an
appropriate dimensional scale factor, a simple combination of the
blade's parameters. The second is a function of the $G_{i}$ only
(for a given shape of the blade), and  must be numerically
evaluated. This means that the solution corresponding to a particular
pair $(G_{x},G_{y})$ can be used to describe a large class of different
physical systems, obtained one from the other by a transformation
of the parameters which leaves $G_{x}$ and $G_{y}$ both unchanged.
For example, the vertical position $Y$ of the blade's tip can be written
as
\begin{equation}
Y=L\,\int_{0}^{1}\sin\theta(\xi)\, d\xi=L\, y(G_{i})\,.\label{eq:cY}\end{equation}
 Scaling the blade length $L$ by a factor $\lambda$, and multiplying
the blade's thickness $d$ by a factor $\lambda^{2/3}$ , we keep
the working point unchanged. This is very useful in order to plan
the integration of the blade into a new system {[}Bertolini 2003{]}.

\subsection{\label{sub:Integration}Integration of the Equation of Statics and
General Properties }

Computing the variation $\delta\tilde{U}$ of the action~(\ref{eq:adimUdef})
with respect to $\theta$ to minimize the potential energy, we obtain
the equation of statics for the physical system in the presence of
an assigned external load $G_{i}$. Then, rewriting $\delta U$ as
a system of first order differential equations we get\begin{eqnarray}
\frac{dp}{d\xi} & = & G_{x}\:\sin\theta(\xi)-G_{y}\:\cos\theta(\xi),\label{eq:basep}\\
\frac{d\theta}{d\xi} & = & \gamma(\xi)\, p,\label{eq:baseq}\end{eqnarray}
with boundary conditions\begin{equation}
\theta(0)=\theta_{0},\qquad\theta(1)=\theta_{1}\label{eq:bc}\end{equation}

\begin{figure}[htbp]
\begin{center}\includegraphics[%
  clip,
  width=1.0\linewidth,
  keepaspectratio]{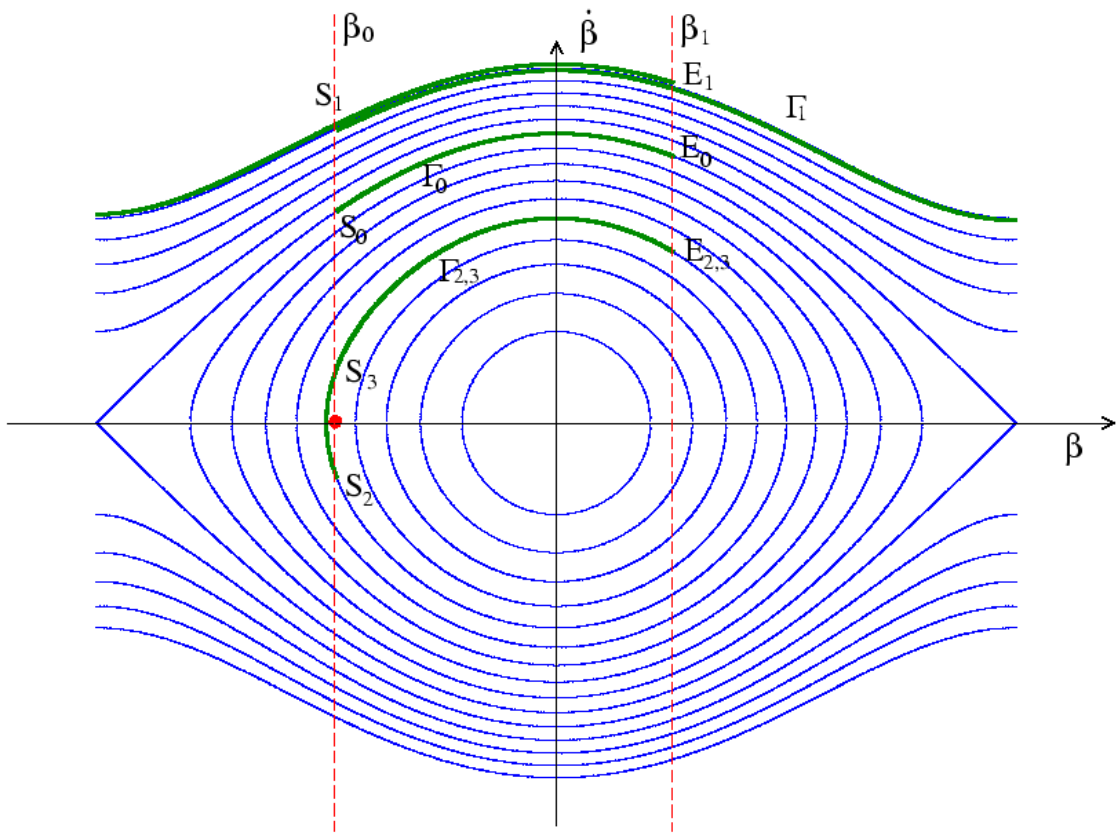}\end{center}

\caption{\label{fig:phase-space}Evolution of the blade's angle in the phase
space, for a blade of constant width. The phase space is equivalent
to the one of the simple pendulum: closed trajectories correspond
to a bounded variation of the blade's angle, open ones to blade's
configurations with turns. Different solutions which satisfy the boundary
condition for $\beta(0)=\beta_{0}$ and $\beta(\sqrt{G})=\beta_{1}$
are depicted: solution $\Gamma_{i}$ starts in $S_{i}$ and ends in
$E_{i}$. We assume periodic boundaries in $\beta$: $\Gamma_{1}$
starts in $S_{1}$ and after a complete $2\pi$ cycle moves to $E_{1}$.
See Section~\ref{sub:Integration} and~\ref{sec:Exact-solution}
for a full discussion.}
\end{figure}

Because this is a boundary value problem, in general the solution
uniqueness cannot be guaranteed. Different equilibrium configurations
can coexist for the same values of boundary conditions and parameters.
This can be easily seen in the particular case of constant blade width.
In this case, $\gamma=1$ and the system admits a $\xi$-independent
quantity $A$ that can be written as

\begin{equation}
A=\frac{1}{2}\left(\frac{d\beta}{d\tau}\right)^{2}+(1-\cos\beta)\label{eq:Pendulum}
\end{equation}
where we set $\beta=\theta-\psi$, $G=\sqrt{G_{x}^{2}+G_{y}^{2}}$,
$\tau=\sqrt{G}\xi$ and $\tan\psi=G_{y}/G_{x}$. This is formally
equivalent to the energy of a simple pendulum and we will discuss now the behavior
of our system from a qualitative point of view, looking at the corresponding
phase diagram depicted in Figure~\ref{fig:phase-space}. In~\ref{sec:Exact-solution}
we give some details about the analytical solution of this problem.

Our formal ``pendulum'' is allowed to evolve only in a well defined
interval of the ``time'' parameter, that is $\tau\in[0,\sqrt{G}]$.
During this interval the system must move from a starting point on the $\beta=\beta_{0}$
line to a final point with $\beta=\beta_{1}$. We can freely choose
our starting point on the line $\beta=\beta_{0}$. Different possibilities
correspond of course to different values of $A$.

Suppose that we start on the boundary of the oscillatory region (the
$S_{0}$ point in Figure~\ref{fig:phase-space}). Then, we find the first
solution $\Gamma_{0}$ when $G$ is large enough to allow the end
point to be in $E_{0}$. This is a very simple configuration, with
a monotonically increasing value of $\theta$. If we now keep $G$ constant
and increase $p$ at the starting point we obtain a motion which is
faster and faster, and a longer and longer trajectory. When $p$ is
big enough, the trajectory will end at $\beta=\beta_{1}+2\pi$, describing
a new solution $\Gamma_{1}$ with the same parameter $G$ as the old
one. Clearly, with this method we can generate an infinite set of solutions
$\Gamma_{k}$, which differ in the number of times the blade's angle
makes a complete turn. Configurations with turns can potentially
give good working points (see for example~{[}Winterflood and Blair
1998{]}), but we are not interested in these in this work.

Another possibility is the existence of {}``undulated'' configurations.
Inside the oscillatory region of the phase diagram we can construct
solutions with the same (large enough) value of $G$, which can be
classified according to the number of zero curvature points (crossings
of the $\dot{\beta}=0$ axis in the phase diagram). To make this plausible
we can consider a solution with some starting point $S_{3}$ (see
Figure~\ref{fig:phase-space}) and a very large $G$, such that it
makes $N$ turns in the phase diagram. When we move $S_{3}$ towards
the boundary of the oscillatory region, we should find at least $2N$
solutions. The reason is that on the boundary the solution is without
crossings, because the time required to approach the point $\dot{\beta}=0$
is infinite. So the trajectory must {}``unwind'' when we move $S_{3}$,
and its end point must cross the $\beta=\beta_{1}$ line at least
$2N$ times. Connected with the existence of {}``undulated'' configurations
is the possibility of buckling-type phenomena%
\footnote{See~{[}Winterflood \etal 2002{]} for a different approach to vertical
seismic attenuation, based on a buckling behavior.%
}. As we will see, this is in fact the key to understanding the behavior
of our system, as will be discussed extensively in the following section.

\subsubsection{\label{sub:Effective-potential}Effective Potential}

To study the solutions of Equations~(\ref{eq:basep})
and (\ref{eq:baseq}) as a function of the parameters, it is convenient
to apply the so-called continuation method. We use the AUTO2000
code, which implements several algorithms for continuation of differential
and algebraic problems. Details can be found in the user's manual~{[}Doedel
\etal 2000{]}.

The basic strategy consists in the exploration of the solution space when smoothly
changing the values of the coordinates of the blade's tip,
the horizontal position
\begin{equation}
x\equiv\frac{X}{L}=\int_{0}^{1}\cos\theta(\xi)\, d\xi\label{eq:cX}
\end{equation}
or the vertical position defined in Eq.~(\ref{eq:cY}).

\begin{figure}[htbp]
\begin{center}\includegraphics[%
  width=1.0\linewidth]{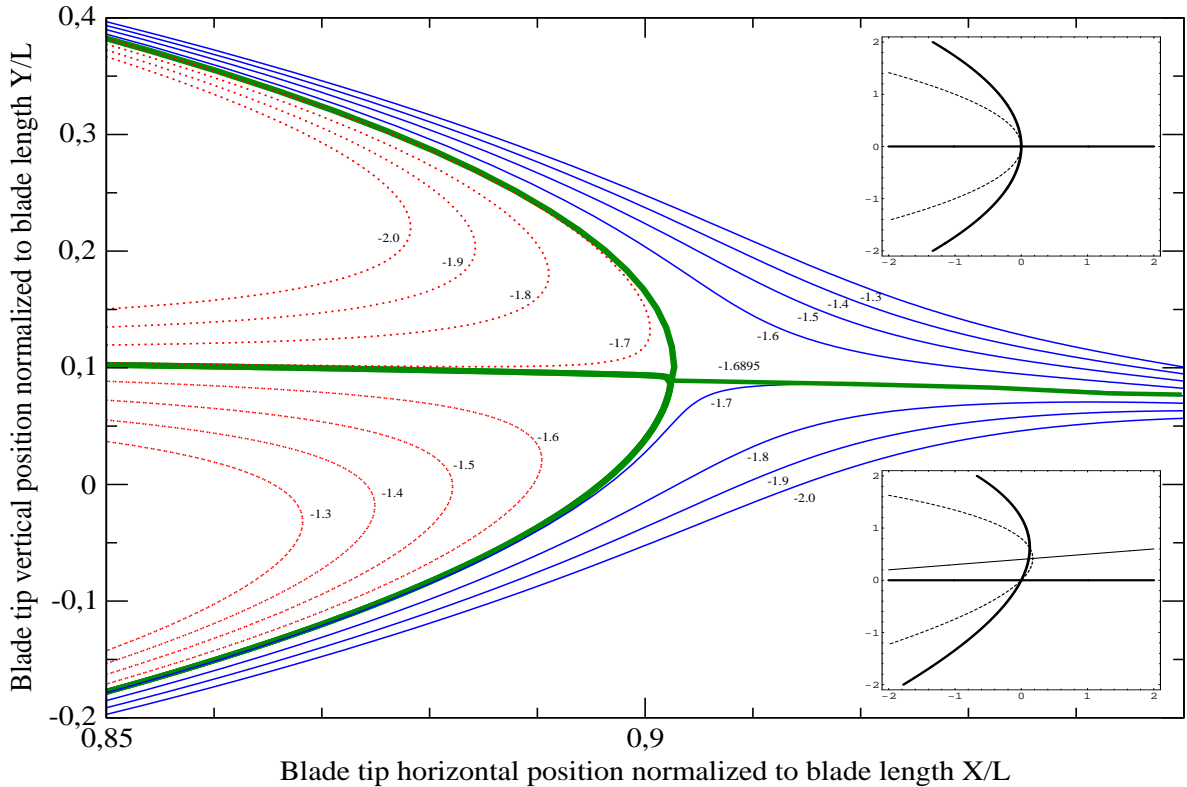}\end{center}

\caption{\label{fig:XY}The normalized vertical position $y$ versus normalized
horizontal position $x$ of the blade's tip, for different values
of the vertical dimensionless load $G_{y}$. For a generic value $G_{y}$
there is a folding singularity where a couple of solutions vanish,
and a stable branch. In the special case $G_{y}=G_{y}^{*}$ (here
$G_{y}^{*}\simeq1.6895$) we obtain a two sided bifurcation with a
turning point. This is depicted schematically in the lower square
(bold lines), emphasizing its difference from the simpler pitchfork
singularity depicted in the upper square. See Subsection~\prettyref{sub:universality}
for a full discussion of these diagrams.}
\end{figure}

Using the continuation method we constructed the plot in Figure~\ref{fig:XY},
which is  a graphic representation of $G_{y}(x,y)$. In the explored region, it appears
that this is a well defined single-valued function as suggested by physical intuition.
In fact, numerical investigation using AUTO2000 gives no evidence of bifurcations. We can repeat a similar study to obtain a representation
of the function $G_{x}(x,y)$ in the same region. We do not show the
relative diagrams because they do not present peculiar features, but
the final conclusions are that both $G_{x}$ and $G_{y}$ are regular
and single-valued functions of the blade's tip coordinate. We can
expand them in small variations of the parameters around the chosen
working point
\begin{equation}
G_{i}(x,y)=G_{i}^{(0)}+K_{ij}\delta_{j}+\frac{1}{2}K_{ijk}\delta_{j}\delta_{k}+\frac{1}{3!}K_{ijkl}\delta_{j}\delta_{k}\delta_{l}+O(\delta_{i}^{4})\label{eq:ExpansionGi}
\end{equation}
where $\delta_{x}=x-x_{0}$ and $\delta_{y}=y-y_{0}$. As $G_{i}$
are the components of the external force acting on the system, it
is natural to ask if they can be written as the partial derivatives
of an effective potential function
\begin{equation}
G_{i}(x,y)=\frac{\partial U_{eff}(x,y)}{\partial x_{i}}.\label{eq:GfromU}\end{equation}
On physical grounds, the answer should be positive.
In fact, this potential energy represents the energy stored in the
system by the external work, and because there is no dissipation in
our model, it should be a well defined function of the position.
If $U_{eff}$ were multivalued, the difference between two of the
energy values could only be interpreted as the difference between
the energy stored in two different configurations with the same $x$,$y$
values. But in this case a closed path in the $(x,y)$ space crossing
a bifurcation point should exist. If $U_{eff}$ exists the integrability
conditions tell us that the coefficients $K_{\cdots}$ are completely
symmetric in their indices.

A comment about the general validity of these results is in order.
As we discussed, it is possible that many solutions with the same
values of $x$ and $y$ exist, so Equations~(\ref{eq:ExpansionGi})
should be understood as a perturbative expansion around some well
defined blade configuration. In practice, this all we need. With
this restriction all should work well, unless we try to expand around
a bifurcation point (that is, a point where the number of solutions
changes).

All the quantities we are interested in can be recovered from the
coefficients $K_{\cdots}$, which are partial derivatives of the effective
potential evaluated at the working point. It is quite important to
find a good method to monitor these quantities during the continuation
procedure. This can be done by introducing auxiliary functions, governed
by appropriate differential equations that must be added to the basic
relations~(\ref{eq:basep}), (\ref{eq:baseq}) and (\ref{eq:bc}). The
method is explained in detail in~\ref{sec: appA}: here we stress
only that for a given order in the expansion of the effective potential
we obtain a finite system of differential equations to solve.

As a final observation, we recall that the assumption of the existence
of an effective potential is not mandatory in order to calculate the
coefficients $K_{\cdots}$. In fact, the integrability conditions
translate into a set of nontrivial relations between the auxiliary
functions which can be used to test or simplify the numerical code
(see~\ref{sec: appA}).

We will give other details about the continuation method in  Subsections~\ref{sub:Vertical-position-of-blade}
and \ref{sub:Vertical-Resonance-Frequency}, explaining its application
to the search for configuration with a low vertical stiffness. For
more information, good starting points are~{[}Doedel \etal 1991,
Doedel \etal 1991b{]} and references therein.

\subsubsection{\label{sub:Vertical-position-of-blade}Vertical Position of the Blade's
Tip}

To give a simple example of application of the continuation method,
we will study the normalized blade tip's vertical position.

Here and in the following we will present results for a particular
configuration with $\theta_{0}=\pi/4$, $\theta_{L}=-\pi/6$ and $\gamma^{-1}(\xi)=c_{1}+c_{2}\cos\beta\xi+c_{3}\sin\beta\xi$,
with $c_{1}=-0.377$, $c_{2}=1.377$, $c_{3}=0.195$ and $\beta=1.361$.
The particular shape of the blade was chosen to compare with measurements
taken on a prototype which will be published elsewhere~{[}Sannibale
\etal 2004{]}. In Subsection~\ref{sub:universality} we will comment
on the general validity of our conclusions.

\begin{figure}[htbp]
\begin{center}\includegraphics[%
  width=0.80\linewidth]{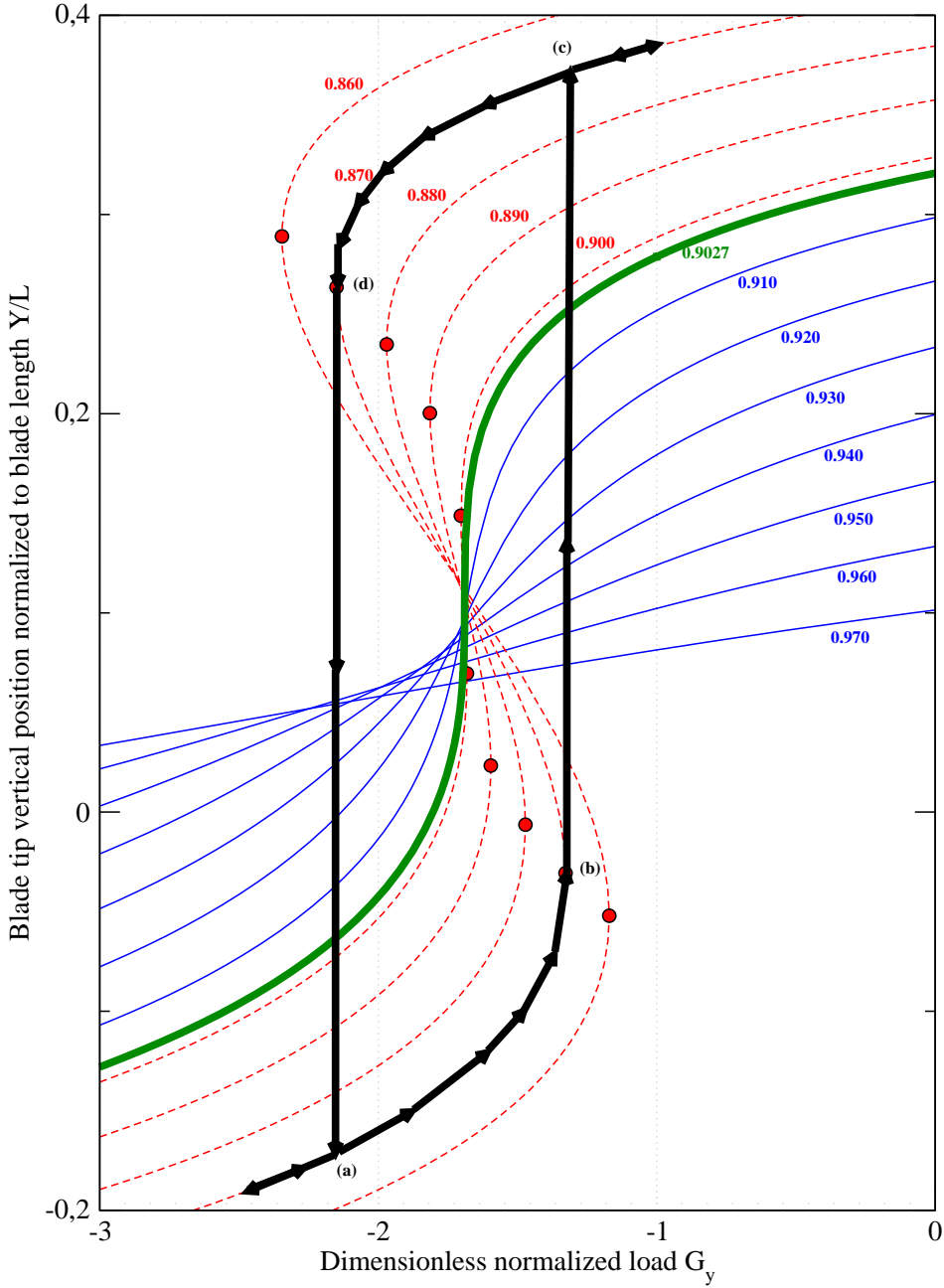}\end{center}

\caption{\label{fig:YV} The normalized vertical position $y$ of the blade's
tip versus the dimensionless vertical load parameter $G_{y}$ (see
Eq.~\ref{eq:Gi}). Each curve corresponds to a different fixed horizontal
position $x$ of the blade's tip (the $x$ value is shown above each
single curve). Two different classes of curves can be distinguished.
For a curve in the first class (depicted with a continuous line) the
vertical position is always an increasing function of the load, so
that each point defines a stable equilibrium point for the system.
For a curve in the second class (depicted with a dashed line) there
is also a region, where $y$ decreases with $G_{y}$, of unstable equilibrium
points. This unstable region is separated from the stable one by a couple of
folding singularities. The continous folded curve, which corresponds
to $x=0.9027$ in this example, separates the two regions. The oriented
closed path shows a typical hysteresis path, and is discussed in Subsection~\ref{sub:Vertical-position-of-blade}.
Note that the two stable equilibrium positions in the bistable region
correspond to different vertical positions of the blade's tip. Boundary
conditions are always $\theta_{0}=\pi/4$ and $\theta_{L}=-\pi/6$.}
\end{figure}

If we choose as independent parameters the vertical load $G_{y}$
and the tip's horizontal position $x$, we can construct the diagram
in Figure~\ref{fig:YV}. There the possible equilibrium values of
$y$ are plotted as a function of the vertical load, for different
values of $x$, in an {}``interesting'' region.

It is evident that the curves can be grouped in two different families,
one made of monotonically increasing functions of $G_{y}$ (continuous
line curves, corresponding to large values of $x$) and the other
made of curves showing a couple of folding singularities (dashed line
curves, folds marked with circles). The boundary between these two
different families corresponds to $x=x^{*}\simeq0.9027$.

A folding singularity can be detected during the continuation procedure
performed by AUTO2000. After the detection it is possible
to add a second free parameter and to continue the fold. This is very
convenient to determine the locus of all the folding singularities.

When $x>x^{*}$ there is only a single equilibrium position $y_{0}$
for each value of the vertical load, and as $\partial y/\partial G_{y}>0$
we can conclude also that the equilibrium is stable. If we adiabatically reduce
$x$, the blade's $y$ will remain at the assigned vertical
load: $y_{0}$ will increase if $G_{y}>G_{y}^{*}=-1.6895$ and it
will decrease otherwise.

When $x<x^{*}$ it becomes possible to have more than one equilibrium
configuration: in that case two new solutions $y_{1}$ and $y_{2}$
appear, and by checking the sign of $\partial y/\partial G_{y}$ it
is easy to see that only the largest and the smallest among $y_{0}$,
$y_{1}$, $y_{2}$ are stable. The system can switch between these
if driven by a large enough external perturbation, and we conclude
that the region $x<x^{*}$ is associated with a bistability. The presence
of this bistability can be described also by observing that the system
can show a kind of hysteresis behavior. This is also illustrated in
Figure~\ref{fig:YV}: suppose we prepare the system in a configuration
which corresponds to the point $(a)$. If we start to increase
the load adiabatically, we observe that the vertical blade tip's position increases
smoothly, moving along the curve, until the fold singularity $(b)$
is reached. At that point the system cannot respond anymore in a smooth
way to a further increase of the load: it will switch abruptly to
the only available equilibrium position $(c)$. Of course the switch
between $(b)$ and $(c)$ points is a dynamical process and depends
on the system inertia. It cannot be described without solving the
nonlinear equation of motion including dissipation mechanisms. Our
static analysis only shows that when the equilibrium is reached the
system must be settled in $(c)$.

If now we reduce the load, the system moves smoothly along the new
curve until it reaches the folding point $(d)$. A further load reduction
triggers a switch to the point $(a)$, completing the cycle.

\subsubsection{\label{sub:Vertical-Resonance-Frequency}Vertical Resonance Frequency}

A very important parameter of an MGASF filter is the angular frequency
$\omega_{y}$ of the fundamental vertical mode. For symmetry reasons
it is evident that during the filter's vertical motion the horizontal
position of the blade's tip is fixed, so the equations which define
our problem are~(\ref{eq:basep}) and (\ref{eq:baseq}), supplemented
by the two fixed boundary conditions~(\ref{eq:bc}) and by the geometrical
constraint~(\ref{eq:cX}). The external parameters are $G_{y}$ and
$x$, while $G_{x}$ is fixed by the constraint.

It is easy to see that in the frequency domain, and for low frequency,

\begin{equation}
\frac{L}{g}\,\omega_{y}^{2}=\frac{1}{F_{y}}\left(\frac{\partial F_{y}}{\partial y}\right)_{x}=\frac{1}{G_{y}}\left(\frac{\partial G_{y}}{\partial y}\right)_{x}=\frac{K_{yy}}{G_{y}}\:.\label{eq:omegadef}\end{equation}

In other words, the MGAS filter is indeed formally equivalent to a simple pendulum
with effective length 
\begin{equation}
L_{eff}=LG_{y}\frac{1}{\left(\frac{\partial G_{y}}{\partial y}\right)_x}.
\label{eq:EffectiveLength}
\end{equation}

For the optimization of filter performance, we are interested in
obtaining the smallest possible value of $\omega_{y}$. As already
stated, this can be obtained by simply increasing $L_{eff}$ while
keeping the working point $G_{i}$ fixed. A possible transformation
that performs this task is a rescaling of the blade length by a factor
$\lambda$\begin{equation}
L\rightarrow\lambda L,\quad\frac{F_{x,y}}{Ed^{3}w(0)}\rightarrow\frac{1}{\lambda^{2}}\:\frac{F_{x,y}}{Ed^{3}w(0)}\label{eq:rescaling}
\end{equation}
which changes the frequency accordingly with
\begin{equation}
\omega_{y}\rightarrow\frac{1}{\sqrt{\lambda}}\:\omega_{y}\:.\label{eq:rescalingW}
\end{equation}
The vertical frequency of the blade decreases only with the square
root of the blade length. A long blade is quite difficult to integrate
into a real system, and its internal resonances start to reduce attenuation
performance at low frequency. For these reasons we are interested
in working points where $L_{eff}$ diverges to infinity.
These are the points of the parameters space where the linearized
response of the system is large, or in other words where the coefficient
$K_{yy}$ of the effective potential goes to zero.

To describe the typical situation we are looking for, let's re-analyze
Figure~\ref{fig:YV}. From Equation~(\ref{eq:EffectiveLength})
it is evident that the effective length of the system is inversely proportional
to the slope of a curve at the considered equilibrium point.
This means that in a folding singularity, $G_{y}=G_{y}^{(fold)}(x)$,
the effective length diverges and so $\omega_{y}\rightarrow0$. Another
point where $L_{eff}\rightarrow\infty$ is on the boundary curve $x=x^{*}$,
for the particular value $G_{y}=G_{y}^{*}$ which corresponds to a
vertical slope.

Setting the parameters of our system {}``near'' one of these critical
points, we can obtain a vertical resonance frequency as small as we
want in principle. However, there is a qualitative difference between
the critical points on a folding singularity $(x,G_{y}^{(fold)}(x))$
and the critical point on the boundary curve $(x^{*},G_{y}^{*})$.

\begin{figure}[t]
\begin{center}\includegraphics[%
  clip,
  width=0.85\textwidth]{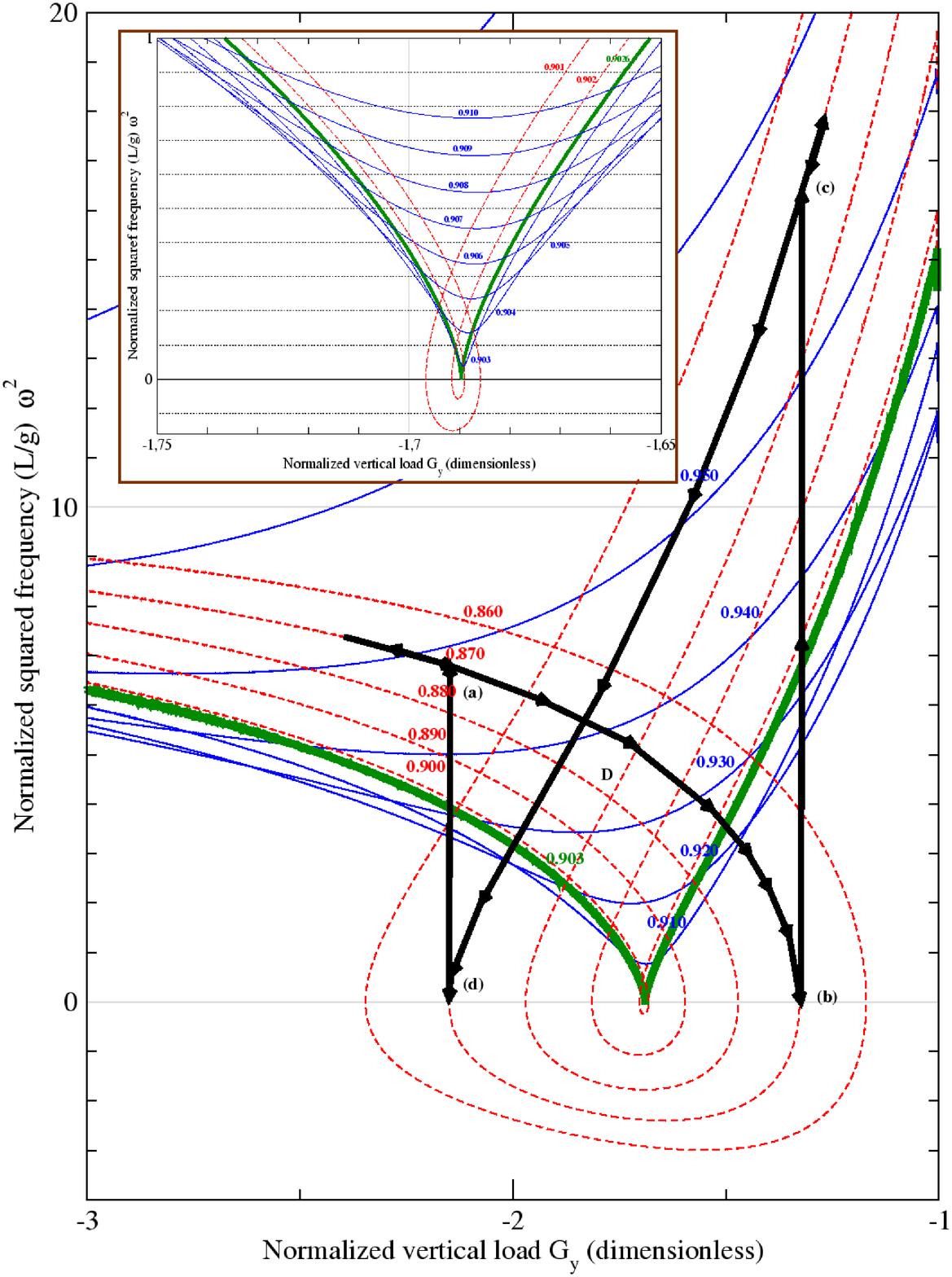}\end{center}

\caption{\label{fig:freqloadZ}Vertical frequency dependence versus load, near
the minimum frequency working point. The vertical axis is the squared
angular frequency of the vertical oscillation, in the low frequency
approximation, expressed in units $g/L$. Each curve is labeled by
its $x$ value. Dashed curves correspond to $x>x^{*}$, continuous
ones to $x<x^{*}$. The bold curve, with a cusp at $\omega_{y}=0$,
is the boundary between the folded and unfolded region. The same hysteresis
cycle of Figure~\ref{fig:YV} is plotted, with the same labels $(a,b,c,d)$.
In the upper left corner we plot a detailed view of the region around
the frequency minimum.}
\end{figure}

This difference can be fully appreciated by looking at Figure~\ref{fig:freqloadZ}.
Here the square of the vertical resonance frequency in $g/L$ units
is plotted versus the vertical load parameter, for different $x$
values. In this picture it is easy to distinguish between the regions
of stable equilibrium points ($\omega_{y}^{2}>0$) and the unstable
one ($\omega_{y}^{2}<0$). An intersection between a curve and the
axis $\omega_{y}^{2}=0$ corresponds to a folding singularity in Figure~\ref{fig:YV}.

We can choose the working point near this intersection, in the stable
region: however in this case a load fluctuation can easily drive the
system into the unstable region, and this is dangerous because we are
in a bistable regime. On the contrary the critical point on the boundary
curve corresponds in the plot to the cusp on the $\omega_{y}^{2}=0$
axis. If we choose our working point near this, on a fully stable
curve $x=x^{*}+\delta_{x}$, $\delta_{x}>0$, it cannot be driven
by a load fluctuation into the unstable region. Of course, a fluctuation
of $x$ (which can be generated by a horizontal motion of the blade's
tip) can drive our system into the unstable region, but in principle
this is less dangerous because we don't have bistability in this region
and the system should fall back to the original point after the fluctuation.
This does not means that there are no problems. The main issue here
is the effect of the nonlinearities that unavoidably appear in systems
with stiffness approaching to zero. This is discussed in section \ref{sub:Nonlinearities}.

We can parameterize Figure~\ref{fig:freqloadZ} with the $K_{\cdots}^{*}$
coefficients. Looking at Figure~\ref{fig:YV} we see that in the
region of interest $G_{y}$ is accurately described by a cubic function
of $y$ with coefficients that depend on $x$. If we take as reference
point $(x^{*},G_{y}^{*})$ and neglect corrections in $\delta_{x}$
we can write

\begin{equation}
G_{y}=\frac{1}{6}K_{yyyy}\delta_{y}^{3}+G_{y}^{*}\label{eq:Gy1}\end{equation}
and for the frequency

\begin{equation}
\frac{L}{g}\omega_{y}^{2}=\frac{\frac{1}{2}K_{yyyy}\delta_{y}^{2}}{G_{y}}\label{eq:freq1}\end{equation}

eliminating $\delta_{y}$ we finally get near the working point

\begin{equation}
\frac{L}{g}\omega^{2}=\frac{K_{yyyy}^{1/3}}{2}\frac{\left[6\left(G_{y}-G_{y}^{*}\right)\right]^{2/3}}{G_{y}^{*}}\label{eq:cusp}
\end{equation}

 where the cusp singularity is evident. Using the same procedure we
can obtain a general relation, more involved, which reproduces quite
accurately Figure~\ref{fig:freqloadZ} for small enough $\delta_{x}\neq0$.

It is interesting to add some considerations about the bistable regime,
drawing the same hysteresis cycle we discussed in Subsection~\ref{sub:Vertical-position-of-blade}.
Looking at Figure~\ref{fig:freqloadZ} it is evident that jumps are
located at the edge of the unstable region, and the two stable equilibrium
points correspond not only to a different blade tip vertical position,
but also to a different value of the vertical resonance frequency.
An exception is, for example, the point $D$.

\subsubsection{\label{sub:Nonlinearities}Nonlinearities}

In Section~\ref{sub:Vertical-Resonance-Frequency} we saw that from
the point of view of system stability the critical point on the boundary
curve $(x^{*},G_{y}^{*})$ should be preferred over the critical point
near a folding singularity. There is another good reason of choosing
$(x^{*},G_{y}^{*})$. The fundamental peculiarity of the MGAS is that
the variation of the stiffness with the position
is in principle not negligible. This means that around
the working point we have to introduce higher order corrections to
the the effective potential. Fixing our attention on the vertical
motion only, and imposing $\delta_{x}=0$, we obtain

\begin{equation}
G_{y}=K_{yy}\delta_{y}+\frac{1}{2}K_{yyy}\delta_{y}^{2}+\frac{1}{6}K_{yyyy}\delta_{y}^{3}+\cdots\label{eq:Gy2}
\end{equation}

which clearly shows the coefficients connected to the anharmonicity.

To quantify anharmonic effects we use the ratio between the first
nonlinear and linear term in the potential

\begin{equation}
\rho_{1}=\sigma\frac{K_{yyy}}{K_{yy}}=\sigma\left(\frac{\partial^{2}G_{y}}{\partial y^{2}}\right)_{x}\left(\frac{\partial G_{y}}{\partial y}\right)_{x}^{-1}\label{eq:rho}
\end{equation}

where the presence $\sigma$, with $\sigma=A/L$, reminds us that
the importance of nonlinearities depends on the amplitude of the motion
$A$. Using Equation~(\ref{eq:omegadef}) and changing the independent
variables to $x$,$G_{y}$ we can rewrite this quantity as

\begin{equation}
\rho_{1}=\sigma\frac{L}{g}\left[\omega_{y}^{2}+G_{y}\left(\frac{\partial\omega_{y}^{2}}{\partial G_{y}}\right)_{x}\right]\label{eq:rhofinal}
\end{equation}

and its behavior for small oscillation frequency can be discussed
looking at Figure~(\ref{fig:freqloadZ}). Near $(x^{*},G_{y}^{*})$
both terms go to zero, so the working point which minimizes the nonlinear
effects is near the minimum of the curve. To first order, we must
set $K_{yyyy}^{*}\delta_{y}+K_{yyyx}^{*}\delta_{x}=0$, where $\delta_{i}$
are the displacements from the critical point and $K_{\cdots}^{*}$
the expansion coefficients of the effective potential around it. Near
a fold, however, only the first term goes to zero, and even worse, the
second term increases without limits. So we expect huge nonlinear
effects near a folding point, and a configuration near $(x^{*},G_{y}^{*})$
should clearly be preferred.

For finite values of frequency the more linear configuration does
not coincide with the configuration that minimizes the frequency variation
with the load. The reason is that a load variation changes both the
working point and the inertia seen by the blade, and only the first
effect is connected to nonlinearities.

We conclude this section with an estimate of the importance of higher
order nonlinearities. The ratio between the second order nonlinear
term in the potential and the linear term gives\begin{equation}
\rho_{2}=\sigma^{2}\frac{K_{yyyy}}{K_{yy}}\simeq\sigma^{2}\frac{K_{yyyy}^{*}}{G_{y}}\left(\frac{g}{L\omega_{y}^{2}}\right)\label{eq:rho2}\end{equation}
which is dangerous because the numerator no longer goes to zero
near the singularity. This result expresses the fact that when the
vertical stiffness of the system goes to zero some kind of nonlinearity
necessarily becomes dominant, and only the cubic term in the effective
potential, proportional to its asymmetry, can be avoided.

This can be seen as a lower limit on the vertical frequency obtainable,
which we expect to be quite small in a concrete situation, because
it is proportional to $\sigma^{2}$:\begin{equation}
f>\frac{\sigma}{2\pi}\sqrt{\frac{g}{L}\frac{K_{yyyy}^{*}}{G_{y}}}\,.\label{eq:freqlim}\end{equation}
As an indication, for a blade with $L=30\, cm$, a coefficient $K_{yyyy}^{*}\simeq10^{3}$
estimated from a typical working point and an indicative amplitude
for the motion $A\simeq10^{-3}\, m$ ,we get $f>10^{-1}\, Hz$, which
seems quite reasonable. The coefficient $K_{yyyy}$ is of course not
the same for all the critical points, so it seems reasonable to find
the best configuration from this point of view, if needed.

\subsubsection{Thermal Stability\label{sec: thermal}}

\begin{figure}[t]
\begin{center}\includegraphics[%
  width=1.0\linewidth]{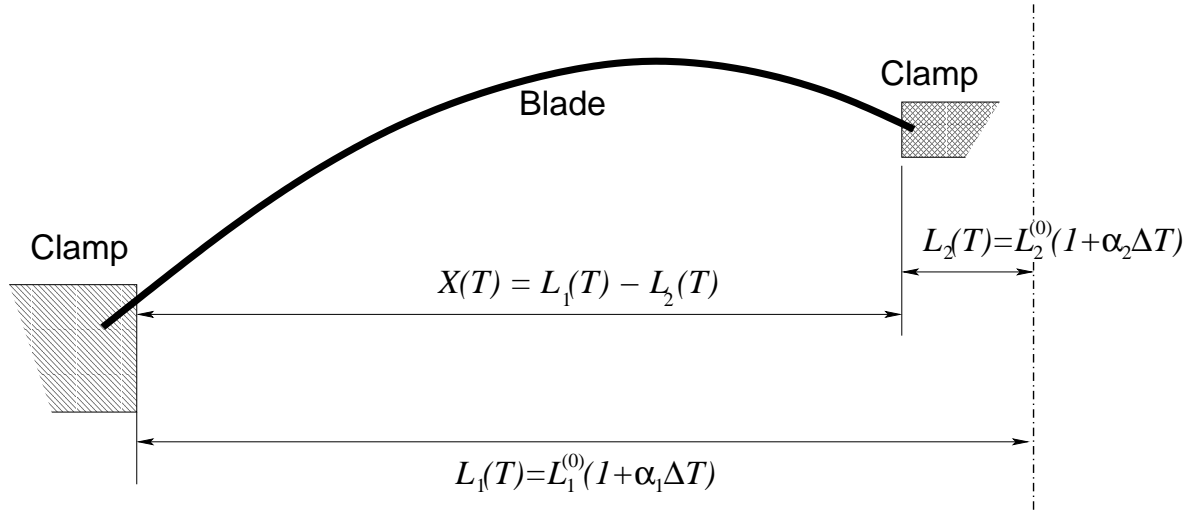}\end{center}

\caption{\label{fig: geotermic}Sketch of geometric blade variations induced
by a temperature change. The angle of the blade's tip is fixed due to the constraint
of the the other blades in the MGAS.
$L_{1}$ and $L_{2}$ are the distances between
each clamp and a fixed vertical reference axis placed in the center
of the filter. Those distances change with the temperature according
to the thermal expansion coefficients $\alpha_{1}$, $\alpha_{2}$
of the clamp materials. $X$ is the horizontal length of the blade. }
\end{figure}

The temperature dependence of the blade equilibrium position depends
on the mechanical and thermodynamical parameters of the material,
i.e. the thermal expansion coefficient and the variation of Young's
modulus with the temperature. A sketch showing how temperature affects
the blade's geometry is shown in Figure~\ref{fig: geotermic}.

Choosing the geometrical and mechanical parameters $L_{1}^{(0)}$,$L_{2}^{(0)}$,$\alpha_{1}$,$\alpha_{2}$
as shown in Figure~\ref{fig: geotermic}, we have that $x=X/L$ changes
linearly with the temperature according to the following law \begin{equation}
x=x_{0}(1+\alpha_{x}\Delta T),\label{eq:xThermalExpansion}\end{equation}
where $\alpha_{x}$, is an effective thermal expansion coefficient
defined by the following formula \begin{equation}
\alpha_{x}=\frac{L_{1}^{0}\alpha_{1}-L_{2}^{0}\alpha_{2}}{L_{1}^{0}-L_{2}^{0}}-\alpha,\label{eq:alphax}\end{equation}
with $\alpha$ the thermal expansion coefficient of the blade's material.
For example, if we choose $\alpha=\alpha_{1}=\alpha_{2}$ (supports
and blade made out of the same material) we obtain $\alpha_{x}=0$.

To first order, the material dependence on temperature enters
through Young's modulus. In fact, we can write\begin{eqnarray}
E & = & E_{0}(1+\alpha_{E}\Delta T),\label{eq:Eexpansion}\end{eqnarray}
where $\alpha_{E}$ is a constant coefficient. Using the definition~(\ref{eq:Gi})
we obtain the effective thermal expansion coefficient for $G_{i}$:

\begin{equation}
\alpha_{G}=-(2\alpha+\alpha_{E}).\label{eq:alphaG}
\end{equation}

The Typical value for stainless steel expansion coefficient at ambient temperature is 
$\alpha \simeq 1 \times10^{-5} K^{-1}$. 
Common values for $\alpha_E$ are negative and of the order $10^{-4} K^{-1}$[articolo NIM]. Accurate values strongly depend on
temperature and on the alloy. 
This means that $\alpha_G$ is positive and dominated by the Young's modulus variation coefficient. 

It is possible  to tune $\alpha_E$ both in amplitude and sign in a
quite large range (say, $\alpha_E \in [-10^{-4} K^{-1},10^{-4} K^{-1}]$) by 
changing alloy composition and/or using thermal treatments [SMC 2003].
Such special alloys can be used indeed, to minimize temperature dependence of the working point.

An interesting question is whether it is possible to find configurations
which are insensitive to the temperature changes for a given set of
coefficients $\alpha_{i}$. As our main issue is to obtain a system
with a small vertical frequency we will limit our study to an understanding
of the thermal behavior of  monolithic blade near a good working
point.

We start with the dependence of the vertical position $Y$ of the
blade's tip as a function of the temperature. The first order variation
of the vertical position due to the temperature is controlled by the
quantity
\begin{equation}
\frac{dY}{dT}=\alpha Y+L\frac{dy}{dT}\label{eq:alphadYdT}
\end{equation}
which as we can see is the sum of two contributions. The first term
accounts for the simple thermal expansion of the blade length. The
second term is the {}``anomalous'' contribution connected to the fact
that the system is displaced from the working point by the temperature
variation. Choosing as independent variables $x$ and $G_{y}$ (see~\ref{sec: appA})
we can write
\begin{equation}
\fl\frac{dY}{dT}=\alpha Y+Lx\left(\frac{\partial y}{\partial x}\right)\alpha_{x}+LG_{y}\left(\frac{\partial y}{\partial G_{y}}\right)\alpha_{G}=L\left(y\alpha+\gamma_{x}\alpha_{x}+\gamma_{G}\alpha_{G}\right)\label{eq:alphay}
\end{equation}

The coefficient $\gamma_{G}$ can be rewritten as\begin{equation}
\gamma_{G}=\frac{g}{L\omega_{y}^{2}}\label{eq:gammaG}\end{equation}
and it is quite dangerous, because it diverges near the critical points.
This is very reasonable physically, because when the vertical stiffness
of the blade becomes very small we can expect a strong sensitivity
to the variation of blade parameters. One possible way to compensate for
this effect is to properly tune the parameters in equation~(\ref{eq:alphay}).
We can adjust the term proportional to $\alpha$ by changing $y$,
which is connected to the boundary conditions for the blade. As we
will see in Subsection~\ref{sub:universality} there is a wide range
of values for the $y$ parameter, positive and negative, compatible
with a critical point.

\begin{figure}
\begin{center}\includegraphics[%
  clip,
  width=1.0\textwidth]{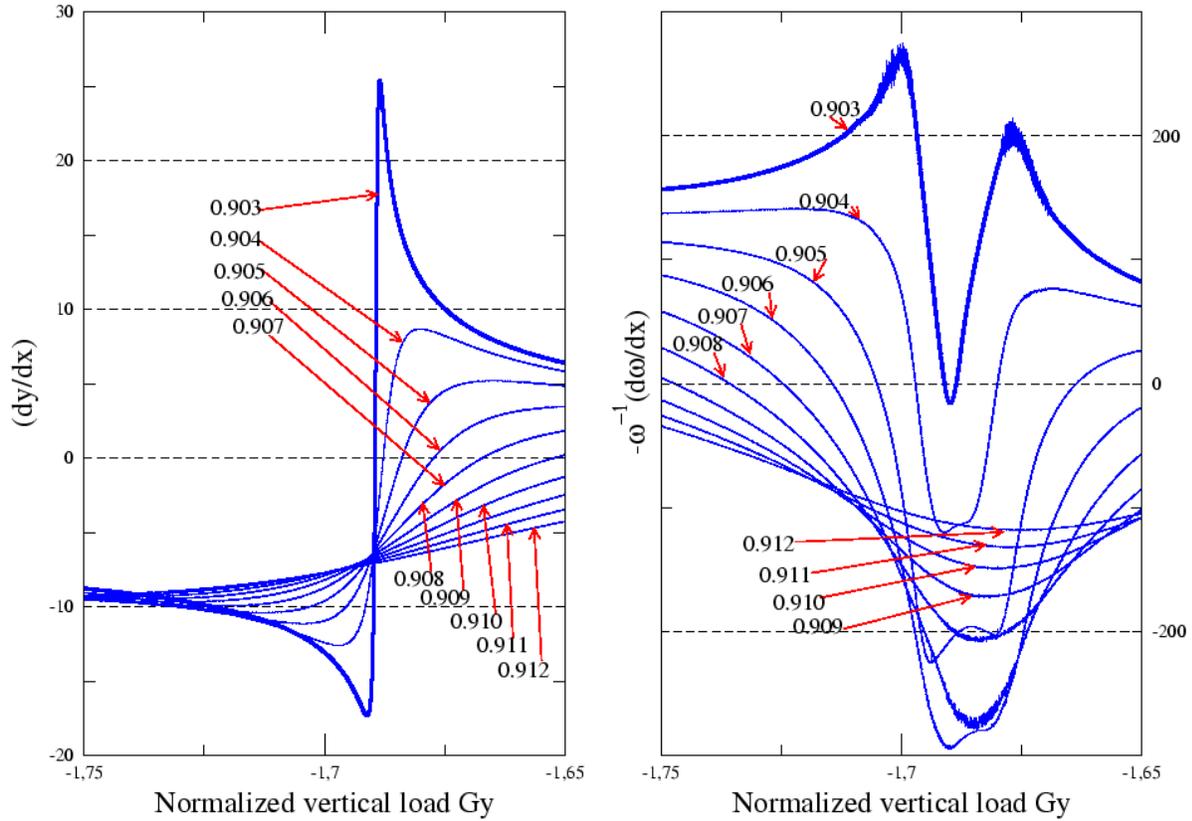}\end{center}

\caption{\label{fig:Thermal}Coefficients related to thermal stability. On
the left side the quantity $(\partial y/\partial x)_{G_{y}}$is plotted
as a function of the load parameter $G_{y}$, for different values
of $x$ choosen near the critical point. On the right side we plot
$\omega_{y}^{-1}(\partial\omega_{y}/\partial_{x})_{G_{y}}$, as a
function of the same parameters of the left side. See Section~\ref{sec: thermal}
for explanations.}
\end{figure}

The term proportional to $\alpha_{x}$, as discussed previously, can
be tuned with an appropriate filter design. We can estimate its
importance by looking at the left diagram in Figure~\ref{fig:Thermal}.
There the relevant quantity $\partial y/\partial x=x^{-1}\gamma_{x}$
is plotted versus $G_{y}$ for different $x$ values in the stable
region near the critical point $(x^{*},G_{y}^{*})$ (compare with
Figure~\ref{fig:freqloadZ}). It is evident that the rapid variation of
this parameter around the critical point is particularly relevant
when we are very near to it ($x\simeq0.903$, bold curve). This means
that it is surely possible to tune this parameter, but also that the
tuning could be very delicate.

Another important parameter, which should be made as much stable as
possible, is the vertical oscillation frequency. We can define an
effective {}``expansion'' coefficient for it in the following way:\begin{equation}
\alpha_{\omega}=\frac{1}{\omega_{y}}\frac{d\omega_{y}}{dT}=\left[\frac{x\alpha_{x}}{\omega_{y}}\left(\frac{\partial\omega_{y}}{\partial x}\right)+\frac{G_{y}\alpha_{G}}{\omega_{y}}\left(\frac{\partial\omega_{y}}{\partial G_{y}}\right)-\alpha\right]\,.\label{eq:alphaW}\end{equation}

There are two quantities which depend on the working point in this
expression. The first one is proportional to $\alpha_{x}$, which
depends on the filter design, and to the variation of the logarithm
of the vertical frequency with $x$, which we can rewrite as\begin{equation}
\frac{1}{\omega_{y}}\frac{\partial\omega_{y}}{\partial x}=-\frac{1}{2}\frac{G_{y}}{\gamma_{G}}\left(\frac{\partial^{2}y}{\partial G_{y}\partial x}\right)\label{eq:THW2}\end{equation}
This contribution depends fully on the nonlinearity of the system.
The quantity~(\ref{eq:THW2}) is plotted on the right side of Figure~\ref{fig:Thermal}
as a function of $G_{y}$, for several $x$ values chosen near the
critical point.

The second working-point-dependent quantity is proportional to the
variation of the frequency with the load. We met this quantity when
we discussed nonlinear effects, and we saw that it can be set to zero
by fixing the working point near one of the minima of Figure~\ref{fig:freqloadZ}.
Here the derivative is weighted with $\omega_{y}^{-1}$, so the situation
is a bit different. As the minima of $\omega_{y}$ are also minima
for $\log\omega_{y}$ it is always possible to set the contribution~(\ref{eq:THW2})
to zero. However, as we approach the critical point its variation will
become very large. As for the vertical variation of the blade's tip,
this means that the tuning of $\alpha_{\omega}$ will be possible
but delicate, expecially very near to the critical point. The method
used to evaluate all of the quantities that appears in these expressions
is explained in~\ref{sec: appA}.

\subsubsection{\label{sub:universality}Universality of the Results}

An important point is the robustness of the critical behavior of a
monolithic blades. We want to understand what is the range for the
{}``general design'' parameters (tip angle, base angle and shape of the blade)
which are compatible with the existence of a critical point.
In order to do that we need a precise  characterization of the
{}``interesting'' working point, that could be used to explore the
parameter space.

This characterization is provided by Figure~\ref{fig:YV}, where
it is apparent that when $(x,y)=(x^{\star},y^{\star})$ we have both\begin{equation}
\left(\frac{\partial G_{y}}{\partial y}\right)(x^{\star},y^{\star})=K_{yy}^{\star}=0\label{eq:ZeroFrequency}\end{equation}
and\begin{equation}
\left(\frac{\partial^{2}G_{y}}{\partial y^{2}}\right)(x^{\star},y^{\star})=K_{yyy}^{\star}=0\,.\label{eq:StableFrequency}\end{equation}
How can these conditions be extracted from ~\prettyref{fig:XY}?
The curves there were obtained by continuing the solution in the $x$
parameter, at a fixed $G_{y}$ and with free $y$. We recognize that
$K_{yy}=0$ when the slope of the depicted curves goes to infinity.
This is a folding singularity that can be identified and continued adding
a new free parameter, for example $G_{y}$, obtaining a locus of points
where the Equation~\prettyref{eq:ZeroFrequency} is satisfied. If,
following this locus, we find another folding singularity, clearly there
also Equation~\prettyref{eq:StableFrequency} becomes true. In this
way the general strategy that can be used to explore the subspace
of interesting working point is clear in principle: add another free
parameter and continue the new fold. Our implementation of the method
was a bit different because with AUTO2000 it is not possible to continue
a fold of a locus of folds. It is based on the use of the auxiliary
functions introduced in~\prettyref{sec: appA}, but in any case give
us equivalent results. For each working point we know the value of
all the parameters, and of course the solution for the equilibrium
equations. Monitoring the auxiliary functions we can also determine
to a given order the effective potential's coefficients $K_{\cdots}$,
to get more quantitative information.

\begin{figure}[htbp]
\begin{center}\includegraphics[%
  bb=0bp 0bp 700bp 800bp,
  clip,
  height=1.0\linewidth,
  keepaspectratio,
  origin=c]{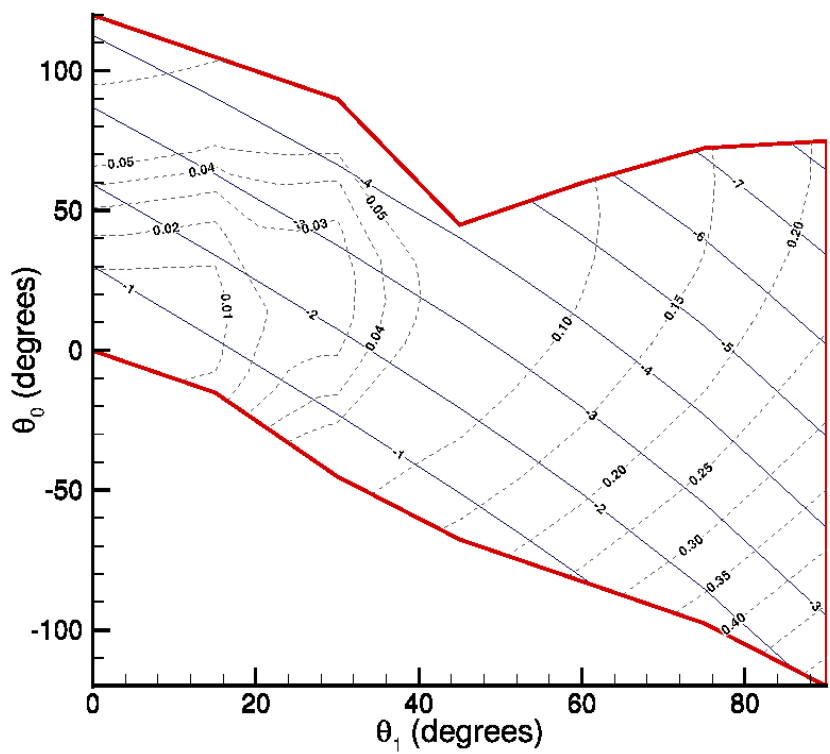}\end{center}

\caption{\label{fig: angle} Some possible working points
for the system. The bottom angle is on the vertical axis, the top
angle on the horizontal one. The set of working points that we found
with the help of the continuation procedure are inside the upper and
lower boundaries which correspond to the bold lines. For a given admissible
pair of boundary angles $\theta_{0}$, $\theta_{1}$ we can read the
value of the vertical load $G_{y}$ (on the continuous contour lines)
and of the $1-x$ parameter (the dotted contour levels). As an example,
it is possible to obtain a good working point for a blade with $\theta_{0}=60$
and $\theta_{1}=0$ by setting $G_{y}=-2$ and $1-x=0.04$.}
\end{figure}

Our preliminary tests show that there is a wide range of parameters
where the MGAS concept is effective. This can be seen
in Figure~\ref{fig: angle}, which can be used to
understand the dependence of the working point on the blade base and
tip angles. For a given pair of angles $\theta_{0}$, $\theta_{1}$
a working point exists only inside the admissible region delimited
by the upper and lower bold boundaries. Inside the admissible region
we can read the values of $G_{y}$ and $1-x$ by following the continuous
and dotted contour lines. It is apparent that a very large range of
possibilities exists, though probably very low values of $G_{y}$
will not really be useful.

Coming back to ~\prettyref{fig:XY} we note that the point
$(x^{*},y^{*})$ where the four bold branches with $G_{y}\simeq-1.6895$
connect together is special. It is clearly a saddle point for $G_{y}$,
and this tells us that the gradient of $G_{y}$ is zero, which means
\begin{equation}
K_{yy}^{\star}=0\label{eq:Kyy}
\end{equation}
and also
\begin{equation}
K_{yx}^{\star}=K_{xy}^{\star}=0\label{eq:Kxy}
\end{equation}

 If the tangent to the curve at that point were vertical we could
classify $(x^{\star},y^{\star})$ as a so called {}``pitchfork''
bifurcation point{[}Iooss and Joseph 1990, Arnold 1996{]}, which is
depicted schematically, with bold lines, in the upper right box. In
the same box the dashed curve represents the locus of points where
$K_{yy}=0$, while the locus of points where $K_{yyy}=0$ is a horizontal
continuous line which overlaps with the bold one. This means that the
{}``interesting'' working point is coincident with the {}``pitchfork''
bifurcation point.

However, a careful inspection reveals that what we really have a
{}``two-sided bifurcation with a turning point''. This one is depicted
schematically in the lower right box of Figure~\ref{fig:XY}, emphasizing
its different character. From this Figure we see that in this case
the intersection between the dashed zero $K_{yy}=0$ curve and the
continuous $K_{yyy}=0$ one does not correspond anymore to the singularity.
More precisely we can expand $G_{y}$ around the crossing point for
the bold lines, obtaining up to the first order in $\delta_{x}$
\begin{equation}
G_{y}-G_{y}^{\star}=\frac{1}{6}K_{yyyy}^{*}\delta_{y}^{3}+\frac{1}{2}\left(K_{yyyx}^{*}\delta_{x}+K_{yyy}^{*}\right)\delta_{y}^{2}+K_{yyx}^{*}\delta_{x}\delta_{y}\label{eq:EffPotSingularity}\end{equation}

where in the pitchfork case (but not in ours) $K_{yyy}^{\star}=0$.
Setting $G_{y}-G_{y}^{\star}=0$ we obtain the equation for the bold
lines in the diagrams, while taking the first and the second derivative
of Eq.~\prettyref{eq:EffPotSingularity} with respect to $y$ we
find the equation for the dashed line\begin{equation}
K_{yy}=\frac{1}{2}K_{yyyy}^{\star}\delta_{y}^{2}+\left(K_{yyyx}^{*}\delta_{x}+K_{yyy}^{*}\right)\delta_{y}+K_{yyx}^{*}\delta_{x}=0\label{eq:Kyyaround}\end{equation}
and for the continuous one\begin{equation}
K_{yyy}=K_{yyyy}^{\star}\delta_{y}+K_{yyyx}^{*}\delta_{x}+K_{yyy}^{*}\label{eq:Kyyyaround}\end{equation}
which of course are valid in a small enough neighborhood around the crossing
point. The main conclusion is that the real singularity associated
with an {}``interesting'' behaviour is a fold of the curve $y(x,\omega_{y}=0)$.
As long as $K_{yyy}$ is small (as it is in our example) we expect
that this singularity will be near the crossing point.

A pitchfork singularity is the one which is usually used to describe
Euler buckling of a beam under compression. In that case the
buckling is associated clearly with the breaking of a symmetry which
is not present in our case. But to completely understand in what sense
the {}``interesting'' working point we described can be associated
to a {}``buckling'' phenomena it would be better to have a connection
with some kind of geometrical peculiarity of the blade's profile associated
with the two different {}``buckled'' configurations. Our conjecture,
supported by some numerical calculations, is that in the bistable regime
the two configurations differ in the presence of a point where the
curvature changes its sign.

\subsubsection{\label{sub:Uniform-Curvature}Uniform Curvature}

At the working point the blade is bent to a profile determined by
Equations~(\ref{eq:basep}) and (\ref{eq:baseq}). We are interested
in configurations with optimally distributed stress. There are two
main reasons for this. First, a blade with a uniform stress has a
higher safety limit. Second, in the optimized case for a given load
the maximal stress is reduced, which means that the intensity of creep
and creak noise is also reduced (special alloys such as properly
treated Maraging do not show an increase of creeping if the stress is increased).

The dominant part of the stress is connected to the curvature radius
$R$ of the blade, so for an optimized configuration this quantity
should be independent of $\xi$. For the {}``old'' GAS filter this
is not possible{[}Bertolini \etal 1999{]} because the boundary conditions
imply zero curvature on the blade's tip. In the monolithic case we
can substitute in the equilibrium equations~(\ref{eq:basep}),(\ref{eq:baseq})
the uniform curvature solution $\theta=\theta_{0}+\xi(\theta_{1}-\theta_{0})$,
obtaining an equation for $\gamma(\xi)$ which can be explicitly solved:

\begin{equation}
\gamma(\xi)=\frac{(\theta_{1}-\theta_{0})^{2}}{(\theta_{1}-\theta_{0})^{2}-G_{y}[\sin\theta(\xi)-\sin\theta_{0}]-G_{x}[\cos\theta(\xi)-\cos\theta_{0}]},\label{eq:ushape}
\end{equation}

where $\gamma(\xi)$ must be always positive in order to describe
a physically acceptable solution.

It is evident that the shape depends explicitly on the $G_{i}$ parameters.
This means that the blade profile should be tailored to the chosen
working point, and it is probably very difficult to do this in practice.
The solution we found tells us only that the uniform curvature solution
is an equilibrium position of the system. To be practically useful
this configuration must be stable, and with a low enough vertical
resonance frequency.

Care must be taken in writing the equations for the continuation procedure.
The dependence of the blade shape from the parameters $G_{i}$ should
be considered in the equations described in~\ref{sec: appA} only
to lowest order. In fact, higher order auxiliary functions $\theta_{(k,l)}$
describe physically the behavior around the working point of a well
defined blade with fixed shape.

We performed extensive numerical explorations of the constant curvature
solution space, trying to find a critical point, but without success.
It seems that the condition of constant curvature is not compatible
with the low vertical stiffness property we want.

After the discussion at the end of Subsection~\prettyref{sub:universality}
this conclusion is very reasonable. We associated the critical point
responsible for the vanishing stiffness with a transition between
two states which differ for the number of points where the curvature
goes to zero. This means that near the critical point a constant curvature
solution should approach a {}``flat'' configuration, but this is
possible only when $\theta_{0}=\theta_{1}$ and the load is parallel
to the blade's direction. This case can be studied analytically, and
is incompatible with a vanishing stiffness.

We take the negative result of our numerical experiments as a confirmation
of our conjecture about the {}``buckling-type'' nature of the critical
point.

\section{Conclusion and Perspectives}

The main purpose of this article was a study of the low frequency
behavior of a geometric antispring. We identified a singularity in
the space of tunable parameters (load and geometrical constraints)
which is associated with a vanishing vertical stiffness.

Nonlinearities do not seem to be big problems for the kind of application
we have in mind, where the ratio between the expected motion's amplitude
and the blade's length $L$ is very small. Cubic nonlinearities can
be eliminated by tuning the working point.

About thermal stability, our conclusion is that it is possible
in principle to reduce the sensitivity of the system to temperature
variations. The tuning can be done by carefully choosing the materials,
the geometry and the working point. However, the quantities involved
typically change very rapidly near the critical point, and this means
that at the end the only possibility will be a case by case trial
and error adjustment procedure, probably delicate and difficult.

We have not yet performed an extensive exploration of the space of
possible configurations. We suspect that this space is very large
indeed, and the possibility of some optimization of the blade's behavior
should not be neglected, if needed. Our next step will be the study
of the behavior of the coefficients $K_{\cdots}^{*}$ which describe
the system around the critical point in this space. We plan also to
study the system in the high frequency regime, including mass effect
and internal modes.

The numerical model described in the article has been implemented
in a set of codes based on the continuation algorithms provided by
the AUTO2000 library. This code will be used in the future to study
and optimize special configurations, and is available on request.

\ack

The authors thanks Alessandro Bertolini for useful information and
discussions about the possibility of the tuning of a material's thermal
properties.

This work was supported in part by MIUR as PRIN MM02248214 and by
the National Science Foundation, United States of America under Cooperative
Agreement No. PHY-9210038 and PHY-0107417.

This contribution has been assigned the LIGO Document number LIGO-PP040004-00-D.

\appendix

\section{First and second order response around the working point\label{sec: appA}}

In principle information about the coefficients of the effective
potential can be recovered by numerical differentiation of the series
$G_{i}(x,y)$ obtained with the continuation method. However, this method
is very inaccurate especially if higher order coefficients are needed.
For this reason we used a different approach to perform this task,
which is basically a perturbative expansion of the external parameters
around a given solution of Equation~\prettyref{eq:basep} and~\prettyref{eq:baseq}.

There are many ways to choose independent parameters, and
here we describe a possibility which is simple from a computational
point of view.

The more direct approach is to choose as independent parameters the
external loads $G_{i}$, and to introduce a small perturbation $\epsilon_{i}$
around a reference point by writing $G_{i}=G_{i}^{(0)}+\epsilon_{i}$.
For a given quantity $C$ we introduce the notation $C_{l,m}$ to
identify the coefficients of its expansion in powers of $\epsilon_{i}$.
For example, we formally write the solutions of the perturbed static
problem as
\begin{equation}
\theta(\xi)=\sum_{k,l}\epsilon_{x}^{k}\epsilon_{y}^{l}\theta_{k,l}(\xi)\,.\label{eq:expansionGG}
\end{equation}
By direct substitution we can easily obtain up to a given order the
equations for $\theta_{k,l}$ and $p_{k,l}$. The equations for
$\theta_{0,0}$ and $p_{0,0}$ are of course exactly the~(\ref{eq:basep})
and~(\ref{eq:baseq}), with the same boundary conditions~(\ref{eq:bc}),
and load parameters $G_{i}=G_{i}^{(0)}$. In the general case we can
write
\begin{eqnarray}
\frac{dp_{k,l}}{d\xi} & = & -\Re\left\{ \left[\left(iG_{x}^{(0)}+G_{y}^{(0)}\right)\eta_{k,l}+i\eta_{k-1,l}+\eta_{k,l-1}\right]e^{i\theta_{0,0}}\right\} \label{eq:ExpansionP}\\
\frac{d\theta_{k,l}}{d\xi} & = & \gamma(\xi)p_{k,l}\label{eq:ExpansionQ}
\end{eqnarray}
where $\eta_{k,l}$ are the expansion coefficient for $e^{-i\theta_{0,0}}e^{i\theta}$
which are
\begin{equation}
\eta_{k,l}=e^{-i\theta_{0,0}}\left.\frac{1}{k!}\frac{1}{l!}\frac{\partial^{k+l}}{\partial\epsilon_{x}^{k}\partial\epsilon_{y}^{l}}e^{i\theta}\right|_{\epsilon_{i}=0}\label{eq:Ekl}\end{equation}
when $k,l\geq0$ and zero otherwise. Note that the equation for $p_{k,l}$
depends linearly on $\theta_{k,l}$ but nonlinearly on $\theta_{k',l'}$
with $k'<k$ or $l'<l$. For example, at the first order we obtain
\begin{eqnarray}
\eta_{1,0} & = & i\theta_{1,0}\label{eq:E10},\\
\eta_{0,1} & = & i\theta_{0,1}\label{eq:E01},
\end{eqnarray}
and at the second order
\begin{eqnarray}
\eta_{2,0} & = & i\theta_{2,0}-\frac{1}{2}\theta_{1,0}^{2},\label{eq:E20}\\
\eta_{1,1} & = & i\theta_{1,1}-\theta_{0,1}\theta_{1,0},\label{eq:E11}\\
\eta_{0.2} & = & i\theta_{0,2}-\frac{1}{2}\theta_{0,1}^{2}.\label{eq:E02}
\end{eqnarray}
We also need  boundary conditions for the newly generated differential equations.
These can be simply obtained by imposing that the angles at the top and at the bottom do not change during the perturbation, obtaining simply
\begin{equation}
\theta_{k,l}(0)=\theta_{k,l}(1)=0\label{eq:expBC}
\end{equation}
for $k>0$ or $l>0$. The coordinates of the blade's tip can also be expressed as a function
of the $\theta_{k,l}$ variables.
By direct substitution in Equations~(\ref{eq:cX}) and~(\ref{eq:cY})
we get \begin{equation}
x+iy=\sum_{k,l}\epsilon_{x}^{k}\epsilon_{y}^{l}\int e^{i\theta_{0,0}}\eta_{k,l}\, d\xi\label{eq:pertXY}\end{equation}
Using these relations we can obtain all the physical predictions we
need. For example, the effective stiffness matrix of the blade's tip
(the $K_{ij}$ piece in the effective potential) is proportional to
the coefficient which connects the first order variation of the coordinate
with the load variation. Explicitly writing first order terms we get:\begin{eqnarray}
(K^{-1})_{xx}=\frac{\partial x}{\partial G_{x}}=x_{1,0} & = & -\int\theta_{1,0}\sin\theta_{0,0}\, d\xi\label{eq:x10}\\
(K^{-1})_{yy}=\frac{\partial y}{\partial G_{y}}=y_{0,1} & = & \int\theta_{0,1}\cos\theta_{0,0}\, d\xi\label{eq:y01}\\
(K^{-1})_{xy}=\frac{\partial x}{\partial G_{y}}=x_{0,1} & = & -\int\theta_{0,1}\sin\theta_{0,0}\, d\xi\label{eq:x01}\\
(K^{-1})_{yx}=\frac{\partial y}{\partial G_{x}}=y_{1,0} & = & \int\theta_{1,0}\cos\theta_{0,0}\, d\xi\label{eq:y10}\end{eqnarray}
 The symmetry in the indices is not apparent, but it can be demostrated,
for example, assuming that the equations for $\theta_{k,l}$ up
to a given order can be obtained by minimizing a potential energy.

An alternative useful expansion can be obtained by perturbing
$G_{y}=G_{y}^{(0)}+\epsilon_{y}$
with fixed $x$. Now the expansion of a generic quantity is
\begin{equation}
C=\sum_{l}\epsilon_{y}^{l}C_{l}\label{eq:altexp}
\end{equation}

and $G_{x}$ is no more an external parameter, but a Lagrange multiplier
which must be determined order by order to satisfy the constraint~(\ref{eq:cX}).

\section{Exact solution for constant width blade\label{sec:Exact-solution}}

If the blade width is constant we can rewrite the potential in Eq.~(\ref{eq:adimUdef}) as

\begin{equation}
\tilde{U}=\int_{0}^{1}\left\{ \frac{1}{2}\left(\frac{d\beta}{d\xi}\right)^{2}-G(1-\cos\beta)\right\} d\xi\label{eq:PendPot}
\end{equation}

where $G=\sqrt{G_{x}^{2}+G_{y}^{2}}$, $\beta=\theta-\psi$ , $\cos\psi=-G_{x}/G$
and $\sin\psi=-G_{y}/G$. This is formally the Lagrangian of a pendulum
which evolves in the {}``time'' $\xi$, whose solution is exactly
known in term of elliptic integrals. We can write the solution as\begin{equation}
\sin\frac{\beta(\xi)}{2}=\mbox{sn}\left(\frac{\sqrt{G}}{k}\xi+\phi,k\right)\label{eq:AnalyticalSol}\end{equation}
where $\mbox{sn}(x,k)$ is a Jacobi elliptic function and $k$,$\phi$
two integration constants that can be determined by imposing the boundary
conditions ($k>0$). It is easy to eliminate $\phi$, and we get the
condition \begin{equation}
\fl\Gamma=\sigma_{1}\mbox{sn}^{-1}\left[\sin\frac{\beta_{1}}{2},k\right]-\sigma_{0}\mbox{sn}^{-1}\left[\sin\frac{\beta_{0}}{2},k\right]-\frac{\sqrt{G}}{k}+(4m+\sigma_{0}-\sigma_{1})K(k)=0\label{eq:BCimpl}\end{equation}
where $m$ is an arbitrary integer, $K(k)$ the complete elliptic
integral of the first kind and $\sigma_{i}=\pm1$. We are interested
in particular in the coordinates of the blade's tip, which can be
written as \begin{equation}
\fl\left(\begin{array}{c}
x(1)\\
y(1)\end{array}\right)=\frac{k}{\sqrt{G}}R(\psi)\left(\begin{array}{c}
\frac{\sqrt{G}}{k}\left(1-\frac{2}{k^{2}}\right)+\frac{2}{k^{2}}\left[f\left(\frac{\beta_{1}}{2},k\right)-f\left(\frac{\beta_{0}}{2},k\right)\right]\\
\sigma_{1}\sqrt{1-k^{2}\sin^{2}\left(\frac{\beta_{1}}{2}\right)}-\sigma_{0}\sqrt{1-k^{2}\sin^{2}\left(\frac{\beta_{0}}{2}\right)}\end{array}\right)\label{eq:Coordites}\end{equation}
where $f(\phi,k)$ is the elliptic integral function of the second kind, and
$R(\psi)$ the matrix which rotates a bidimensional vector by an angle
$\psi$. We also have to take into account that during a variation
of the parameters the boundary conditions stay fixed. This gives us
a relation, which can be obtained by imposing $d\Gamma(1)=0$. Another
constraint is that the coordinate $x$ of the blade's tip does not
change during the variation, which is equivalent to imposing $dx(1)=0$. The
last step is to evaluate the differential $dy(1)$: in this way we
get three linear relations between $dy(1)$, $dG_{x}$, $dG_{y}$ and
$dk$ that we can use to express\begin{equation}
dy(1)=\chi\, dG_{y}\label{eq:Chi}\end{equation}
where $\chi$ is a function of the working point, and $\chi\rightarrow\infty$
when the vertical stiffness of our costrained blade goes to zero.
After some algebraic multiplication it is easy to see that we can
express $\chi$ as
\begin{equation}
\chi=\left|\begin{array}{ccc}
\frac{\partial\Gamma}{\partial G} & \frac{\partial\Gamma}{\partial\psi} & \frac{\partial\Gamma}{\partial k}\\
\frac{\partial x}{\partial G} & \frac{\partial x}{\partial\psi} & \frac{\partial x}{\partial k}\\
\frac{\partial y}{\partial G} & \frac{\partial y}{\partial\psi} & \frac{\partial y}{\partial k}\end{array}\right|\left|\begin{array}{cc}
\frac{\partial\Gamma}{\partial G} & \frac{\partial\Gamma}{\partial\psi}\\
\frac{\partial x}{\partial G} & \frac{\partial x}{\partial\psi}\end{array}\right|^{-1}\label{eq:Chiexplicit}\end{equation}
where all the quantities are evaluated at $\xi=1$. In order for $\chi$
to be divergent there are two possibilities: some of the partial derivatives
inside the determinants could be divergent, or the $2\times2$ determinant
could go to zero. Moreover, in the first case, as the determinant depends
linearly on its elements, the only possibilities are a divergence
of partial derivatives which appears in the $3\times3$ but not in
the $2\times2$ determinant.

We used the analytical solution for cross-checking purposes and as
a convenient starting point for the continuation procedure. By introducing
an interpolation parameter it is easy to continue the blade profile
from the constant width case to the specific case we are interested
in.

\References

\item[] Arnold V I 1996 {\it Geometrical methods in the theory of ordinary differential equations} (Springer Verlag)

\item[] Beccaria M \etal 1997 \NIM A {\bf 394} 397--408

\item[] Bertolini A, Cella G, DeSalvo R and Sannibale V 1998 \NIM A {\bf 435} 475--83

\item[] Bertolini A 2003 {Proceedings $9^{th}$ Pisa meeting on advanced detectors}

\item[] Caron B \etal 1997 \CQG {\bf 39} 1461--9

\item[] Cella G, DeSalvo R, Sannibale V, Tariq H, Viboud N and Takamori A 2002 \NIM A {\bf 487} 652--60

\item[] Doedel E J, Keller H B and Kern�ez J P 1991 {\it Int. J. Bifurcation and Chaos} {\bf 1} 493--520

\item[] \dash 1991b {\it Int. J. Bifurcation and Chaos} {\bf 1} 745--72

\item[] Doedel E J, Paffenroth R C, Champneys A R, Fargrieve T F, Kuznetsov B S and Wang X J 2000 {\it AUTO2000: Continuation and bifurcation software for ordinary differential equations} (Applied and Computational Mathematics, California Institute of Technology, available from http://indy.cs.concordia.ca)

\item[] Iooss G and Joseph D D 1990 {\it Elementary stability and bifurcation theory} (Springer Verlag)

\item[] Landau L D and Lifshitz E M 1959 {\it Theory of Elasticity} (Pergamon Press UK)

\item[] Sannibale V \etal 2004 {\it In Preparation}

\item[] M. Beccaria \etal 1997 {\it Extending the VIRGO gravitational wave detection band down to a few Hz: metal blade springs and magnetic antisprings} \NIM A {\bf 394} 397-408

\item[] SMC 2003 {\it Technical Report Special Metals Corporation SMC-086 } (available from http://www.specialmetals.com)

\item[] Winterflood J, Barber T and Blair D G 2002 \CQG {\bf 19} 1639--45

\item[] Winterflood J and Blair D G 1998 \PL A {\bf 243} 1--6

\endrefs
\end{document}